# Conductive filament evolution dynamics revealed by cryogenic (1.5 K) multilevel switching of CMOS-compatible $Al_2O_3/TiO_2$ resistive memories

Yann Beilliard[1,2,3*], François Paquette[1,2], Frédéric Brousseau[1,2], Serge Ecoffey[1,2], Fabien Alibart[1,2,4], Dominique Drouin[1,2,3]

[1]*Institut Interdisciplinaire d'Innovation Technologique (3IT), Université de Sherbrooke, Sherbrooke J1K 0A5, Canada*
[2]*Laboratoire Nanotechnologies Nanosystèmes (LN2) – CNRS UMI-3463 – 3IT, Sherbrooke J1K 0A5, Canada*
[3]*Institut Quantique (IQ), Université de Sherbrooke, Sherbrooke J1K 2R1, Canada*
[4]*Institute of Electronics, Microelectronics and Nanotechnology (IEMN), Université de Lille, 59650 Villeneuve d'Ascq, France*

*\*Correspondance:* yann.beilliard@usherbrooke.ca – *Orcid:* https://orcid.org/0000-0003-0311-8840

**Abstract** – Non-volatile resistive switching devices are considered as prime candidates for next-generation memory applications operating at room temperature and above, such as resistive random-access memories (RRAM) or brain-inspired in-memory computing. However, their operability in cryogenic conditions remains to be mastered to adopt these devices as building blocks enabling large-scale quantum technologies via quantum-classical electronics co-integration. This study demonstrates multilevel switching at 1.5 K of $Al_2O_3/TiO_{2-x}$ resistive memory devices fabricated with CMOS-compatible processes and materials. The *I-V* characteristics exhibit a negative differential resistance (NDR) effect due to a Joule-heating-induced metal-insulator transition of the $Ti_4O_7$ conductive filament. Carrier transport analysis of all multilevel switching *I-V* curves show that while the insulating regime follows the space charge limited current (SCLC) model for all resistance states, the conduction in the metallic regime is dominated by SCLC and trap-assisted tunneling (TAT) for low- and high-resistance states respectively. A non-monotonic conductance evolution is observed in the insulating regime, as opposed to the continuous and gradual conductance increase and decrease obtained in the metallic regime during the multilevel SET and RESET operations. Cryogenic transport analysis coupled to an analytical model accounting for the metal-insulator-transition-induced NDR effects and the resistance states of the device provide new insights on the conductive filament evolution dynamics and resistive switching mechanisms. Our findings suggest that the non-monotonic conductance evolution in the insulating regime is due to the combined effects of longitudinal and radial variations of the $Ti_4O_7$ conductive filament during the switching. This behavior results from the interplay between temperature- and field-dependent geometrical and physical characteristics of the filament.

The ever-increasing improvement of nanofabrication processes has led to promising demonstration of high-quality solid-state quantum devices functioning at sub-Kelvin temperatures [1,2]. Efforts are currently focused towards the development of silicon-based quantum systems that could be operated above 1 K [3,4]. These works pave the way to truly scalable quantum bit architectures leveraging mature CMOS technologies [5]. However, current classical electronic tools and rack-scale instruments used to control few-qubit systems from outside the cryostat cause major scalability, automation and performance issues hindering the fabrication of full-scale quantum computers. Overcoming these roadblocks therefore requires fully integrated cryogenic quantum-classical interfaces able to control qubits from inside refrigerators. Such interface would be composed of CMOS-based digital and analog electronics, enabling mixed-signal control systems [6–10]. In that scope, non-volatile cryogenic memory technologies hold a key role to store data related to qubit states and error correction algorithms. While standard DRAM-based cryogenic systems have been investigated at 77 K [11,12], emergent nanoscale resistive memory devices with better scalability could play a key role to enable large-scale quantum systems.

Resistive random-access memory (RRAM) based on crossbar arrays of non-volatile resistive memories are considered as prime candidates for next generation high-density data storage applications [13,14]. Device quasi-static and dynamic non-linearity, multi-bit programing precision and variability are all examples of challenges preventing large-scale adoption of RRAM technology [15]. Additionally, cryogenic applications such as quantum-classical interfaces require low electrical noise and power dissipation. Fundamental investigations have to be conducted on resistive memories in cryogenic conditions to both demonstrate reversible, non-volatile and multilevel resistive switching and to better understand the conductive filament geometry evolution during switching. Studies on temperature-dependant behavior and conduction mechanisms of devices based on transition metal oxides have been reported for temperatures down to 4 K [16–24]. In addition, we previously reported successful binary resistive switching of $Al_2O_3/TiO_{2-x}$ devices at 1.5 K [25]. At that temperature, $TiO_2$-based memories exhibits negative differential resistance (NDR) effects related to a Joule-heating-induced metal-insulator transition (MIT) of the $Ti_4O_7$ conductive filament (CF) [19,20]. The resulting highly non-linear current-voltage characteristics could be leveraged for the implementation of selector-free multibit memory crossbar arrays [26]. It is however necessary to better understand the behavior of TiO$_2$-based resistive memories in cryogenic conditions, specifically for multilevel switching.

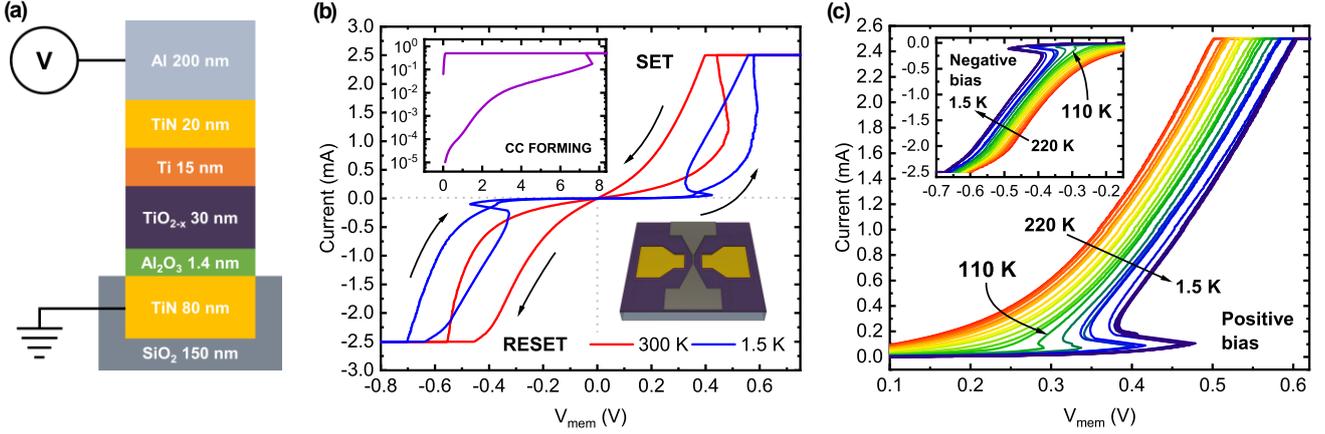

FIG. 1. (a) Schematic cross-section view of the resistive memory studied. (b) Resistive switching cycles performed at 300 K (red lines) and 1.5 K (blue lines). The top-left inset shows the forming process with a current compliance of $I_c = 500$ µA. The schematics in the bottom-right corner illustrate the crosspoint structure. (c) I-V curves for LRS in function of temperature in the range 220-1.5 K for positive bias. Negative bias measurements are shown in inset. A complete switching was performed at each temperature stage. The NDR effect associated to the MIT starts to appear at a critical cryostat temperature of $T_c \approx 110$ K.

In this work, we report and investigate multilevel resistive switching of TiN/Al$_2$O$_3$/TiO$_{2-x}$/Ti/TiN memory devices conducted at 1.5 K. Full switching cycles are first performed between 300 and 1.5 K, revealing MIT-induced NDR effects below 110 K. Analysis of the current-voltage (I-V) curves are then conducted for the multilevel SET and RESET processes, showing that space-charge limited current (SCLC) and trap-assisted tunneling (TAT) are dominant for specific field ranges and resistance states. While the conductance programming of the device is continuous and gradual in the metallic regime (MR), the insulating regime (IR) is characterized by a non-monotonic conductance evolution. This behavior is further investigated thanks to an analytical model accounting for transport mechanisms, resistance states and NDR effects, all based on a two-phase switching-dependent variable geometry cylindrical CF.

The studied devices are resistive memory crosspoints with a symmetrical 2 µm wide TiN electrode architecture and a switching stack made of Al$_2$O$_3$/TiO$_{2-x}$. The fabrication process was conducted on thermally oxidized 1×1 cm² silicon substrates. All lithography steps were done using contact UV lithography. The 80 nm thick TiN bottom electrodes (BE) were fabricated with a CMOS-compatible damascene process involving UV lithography, plasma etching of SiO$_2$, TiN sputter deposition and chemical-mechanical planarization (CMP). The switching stack composed of a 1.4 nm thick layer of Al$_2$O$_3$ and a 30 nm thick layer of non-stoichiometric TiO$_{2-x}$ was then deposited by atomic layer deposition (ALD) and sputtering respectively. Finally, the 20 nm thick TiN and 200 nm Al top electrodes (TE) were fabricated. Figure 1(a) shows a schematic cross-section detailing the thicknesses of all materials.

Electrical characterizations were conducted from 300 to 1.5 K with an Agilent E5270B parametric measurement in a Janis variable temperature insert (VTI) cryostat. For all measurements, the BEs were grounded and the signals were applied to the TEs. Prior to any resistive switching, a current-controlled forming process was conducted with a compliance of $I_c = 500$ µA. The average resistance of the BEs and TEs were measured at each temperature, allowing the calculation of the voltage $V_{mem}$ across the switching region by subtracting to the applied voltage $V_{applied}$ the drop of voltage resulting from access resistances.

Figure 1(b) shows binary resistive switching operations performed at 300 and 1.5 K, with the current plotted against $V_{mem}$ data. The I-V characteristics at 300 K are typical of TiO$_2$-based devices with gradual SET/RESET processes [27]. The highly nonlinear curves and NDR effects observable at 1.5 K are caused by the Joule-heating-induced MIT of the Magnéli phase Ti$_4$O$_7$ inside the CF [19,25], known to occur in the temperature range of 120-155 K [28,29]. The temperature-dependent I-V curves of a device in low-resistance state (LRS) depicted in Fig. 1(c) allow to pinpoint the critical cryostat temperature of $T_c \approx 110$ K for both positive and negative bias (see inset). The threshold voltage $V_{th}$ corresponding to the onset of the MIT-based NDR increases with decreasing temperature, due to the higher power needed to heat up the CF up to the temperature range of 120-155 K.

Multilevel resistive switching operations are then successfully performed at 1.5 K, as shown in Fig. 2(a). The multilevel SET process is achieved with current-controlled incremental switching where $I_{applied}$ goes from 0.8 mA to 2.5 mA with a step of $\Delta I_{SET} = 0.1$ mA. The multilevel RESET procedure is then conducted by voltage-controlled incremental switching with $V_{applied}$ comprised between -2.5 V resistance state (IRS) in the IR (point A), while an IRS in the MR corresponds to the LRS in the IR (point B). Regarding the RESET process, the LRS and HRS in the MR become respectively the HRS and LRS in the IR (points D and E).

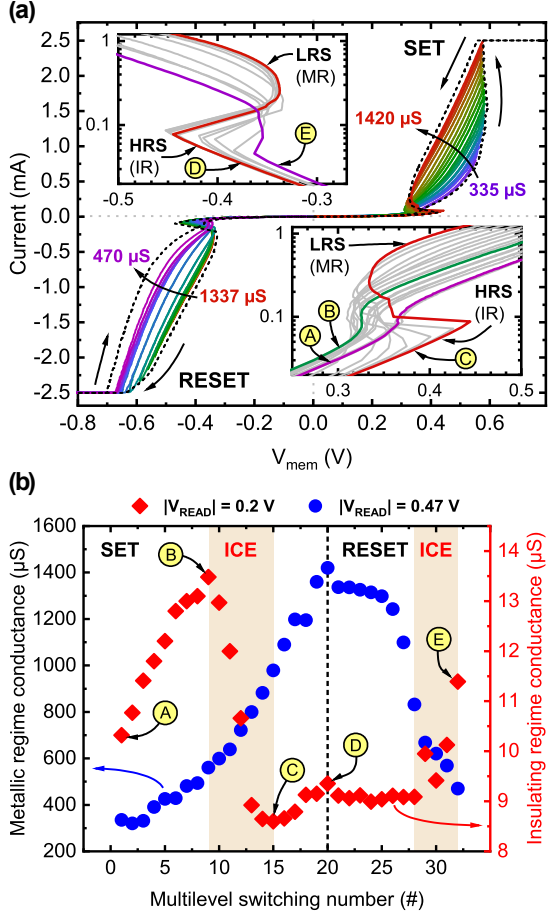

To better illustrate this behavior, Fig. 2(b) shows the differential conductance values in the IR and the MR measured at 0.2 V and 0.47 V respectively for each multilevel switching step. As previously mentioned for $TiO_2$-based memory devices, the multilevel conductance evolution in the MR exhibits a continuous and gradual increase (decrease) during the SET (RESET) process. However, in the IR the conductance exhibits a non-monotonic evolution, where its value first increases from 10.3 µS up to 13.5 µS (points A and B), then decreases down to 8.6 µS revealing the ICE compared to the MR (point C), to finally increase again up to 9.3 µS (point D). Regarding the RESET process, the conductance first decreases slightly down to ~9 µS, only to increase after the 8th RESET operation to reach 11.4 µS (point E). These observations indicate that for a reliable read and usage of the conductance state in the scope of cryogenic high-density memory applications, the read voltage $V_{READ}$ should be carefully chosen to be both in the MR and below the writing voltage $V_{WRITE}$.

In order to better understand the difference in the conductance evolution between the insulating and metallic regimes of the device, we perform a conduction mechanism analysis based on the I-V curves of the multilevel resistive switching at 1.5 K. Similarly to previous transport analysis conducted on $TiO_2$-based devices at low temperature [25,30], we found that space charge limited current (SCLC) and trap-assisted tunneling (TAT) are the dominant conduction mechanisms. Figure 3 shows the results of the $\ln(I)$ vs. $\ln(V)$ and $\ln(J)$ vs. $1/E$ plots, corresponding to the SCLC and TAT models respectively, where oxygen vacancies act as traps and whose current density is described as [31,32]

$$J_{SCLC} = \frac{9}{8}\varepsilon_0\varepsilon_r\mu\theta\frac{V^2}{L^3} \quad (1)$$

and

$$J_{TAT} = A\,exp\left(\frac{-8\pi\sqrt{2qm^*}}{3hE}\Phi_T^{3/2}\right). \quad (2)$$

Here, $\varepsilon_0$ is the vacuum permittivity, $\varepsilon_r$ is the dielectric constant of the switching stack, $\mu$ is the electron mobility, $\theta$ is the ratio of free and shallow trapped charge, $L$ is the thickness of the active stack, $A$ is a constant, $q$ is the electronic charge, $m^*$ is the effective mass of the electron in the oxide (~$9m_0$ for the electrons in $TiO_2$ [33]), $h$ is the Planck's constant, $E$ is the electric field and $\Phi_T$ is the energy of the trap below the oxide conduction band minimum. The linear fits of the $\ln(I)$ vs. $\ln(V)$ plots of the multilevel SET and RESET processes depicted in Figs. 3(a) and 3(b) can be described by $I \propto V^m$, where the index $m$ is the slope of the fit. These results suggest that SCLC is dominant in all resistance states at low and -5 V with a step of $\Delta V_{RESET} = -0.25$ V. This corresponds to a voltage $V_{mem}$ across the device of -0.48 V and -1.61 V respectively. The values of differential conductance $dI/dV$ measured at $|V_{READ}| = 0.47$ V visible in Fig. 2(a) range from 335 µS to 1420 µS during the multilevel SET, then go from 1337 µS to 470 µS during the multilevel RESET. The insets illustrate the I-V curves for the SET and RESET processes in semi-logarithmic scale centered on the NDR effects. One can notice a non-monotonic and partly inverse conduction evolution (ICE) in the IR with respect to the continuous conductance increase in the MR for both SET and RESET procedures. For the SET process, the high-resistance state (HRS) in the MR indeed corresponds to an intermediate-bias (i.e. insulating regime, see insets), but only in high-resistance states at high bias (i.e. metallic regime). In the IR, the slope of $m \approx 2$ indicates that the current level is controlled by the oxygen vacancies acting as traps ($I \propto V^2$). The slope of $m \approx 4$ in the MR indicates that the conduction is governed by the power law $I \propto V^m$, corresponding to the SCLC model where the traps are being filled. According to the ln(*J*) *vs.* 1/*E* plots for the MR shown in Figs. 3(c) and 3(d), the dominant transport mechanism gradually changes from trap-filling SCLC to TAT as the resistance state gradually change from HRS to LRS. This transition could be explained by the increase in length of the conductive filament during the multilevel switching, which seems to remain only partially formed even after the last SET process. As the gap between the tip of the CF and the bottom electrode closes, the trap-controlled transport of the electrons is replaced by a single trap-assisted tunneling event. Using the slope of the fits and the equation (2), trap energy barrier values of around 16 and 21 meV were extracted for IRS and LRS respectively. These values are in agreement with previous studies on similar devices characterized at 1.5 K [25].

FIG. 2. (a) Multilevel SET and RESET processes at 1.5 K. Each color corresponds to a single switching operation. The black dotted lines represent the I-V curves for a full SET/RESET cycle. Insets show the I-V characteristics for SET and RESET processes centered on the NDR effects plotted in semi-logarithmic scale, highlighting the non-monotonic conductance evolution in the IR compared to the MR. (b) Differential conductance values as a function of the multilevel switching number, calculated at $|V_{READ}| = 0.2$ V (IR) and $|V_{READ}| = 0.47$ V (MR). The conductance evolution in the MR is notably more linear during the SET compared to the RESET, mostly due to the current-controlled approach. The number of conductance states achieved for SET and RESET are 20 and 12 respectively. Inverse conductance evolutions (ICE) in the IR with respect to the conductance evolution in the MR are highlighted in yellow for both SET and RESET. In both (a) and (b), circled letters A, B, C, D and E correspond to the different key resistance states in the IR.

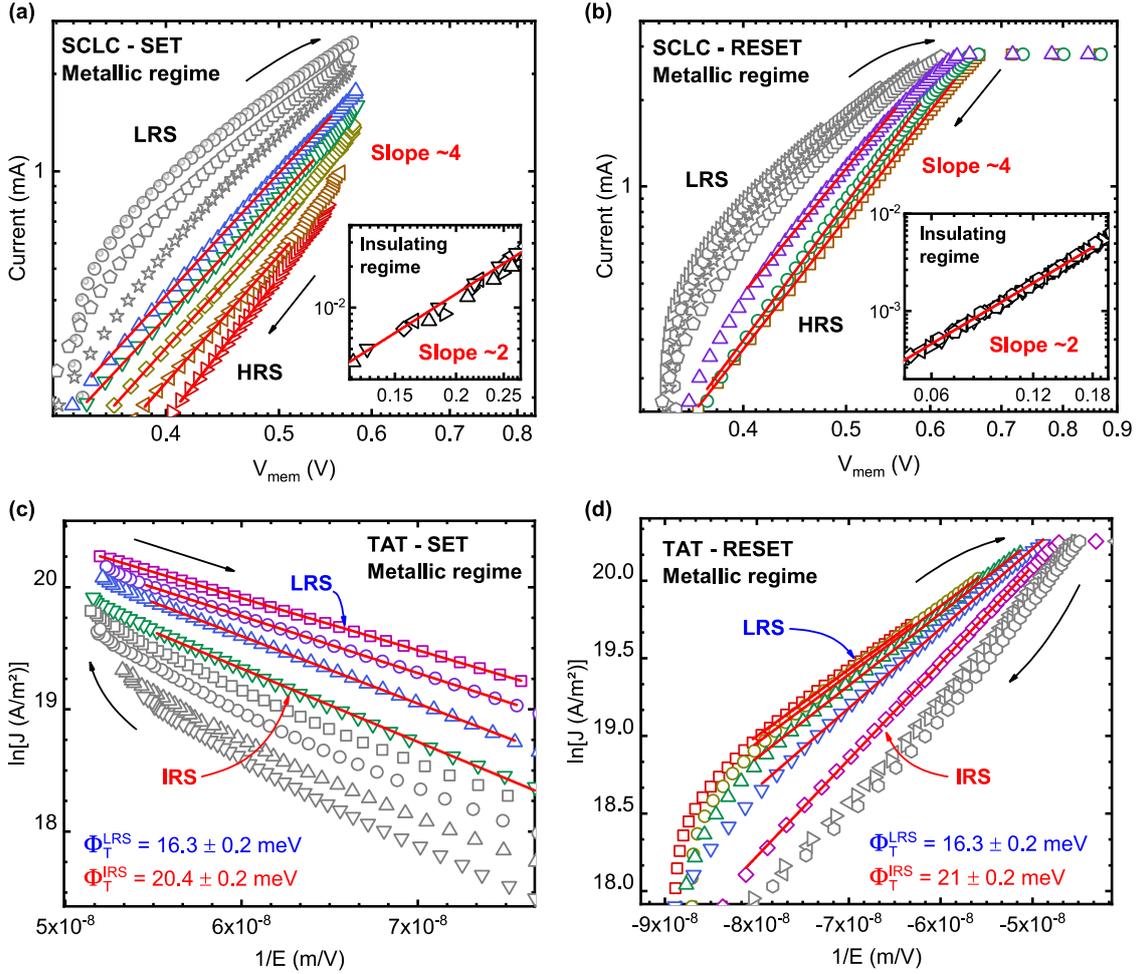

FIG. 3. Conduction mechanism analysis of the multilevel switching curves at 1.5 K. (a, b) ln(*I*) *vs.* ln(*V*) plots for the SET and RESET processes in the voltage range of the MR, with the insets showing the results in the voltage range of the IR. All curves in the IR can be linearly fitted with a slope of ~2, in agreement with the SCLC model. In the MR, only the higher resistance states show linear fits of slope ~4, which suggests a trap-filling SCLC regime. (c, d) TAT plots for the SET and RESET processes showing linear fits for the lower resistance states in the MR only. The extracted trap energy barrier values are given for the IRS and LRS. The colored symbols of the ln(*J*) *vs.* 1/*E* and ln(*I*) *vs.* ln(*V*) plots indicate when the transport behavior best follows the SCLC model (a, b) and the TAT model (c, d).

Following these experimental results, a current-controlled conduction model accounting for the MIT-based NDR effects and the different resistance states was employed in order to gain further insights into the ICE between the IR and MR. This model is based on a 2D analytical approach initially developed by Pickett *et al.* [19]. Figure 4(a) illustrates one half of the system considered, along with the corresponding equivalent circuit diagram. The system is composed of a cylindrically symmetric CF of length $L_{CF}$ and radius $r_{CF}$, separated from the bottom electrode by a gap of length $w$. The MIT contribution to the model comes from the CF being considered as a two-phase channel made of a self-heating metallic core of radius $r_{met}$ and resistivity $\rho_{met}$, surrounded by an insulating shell of radius $r_{ins}$, resistivity $\rho_{ins}$ and thermal conductivity $\kappa$. The normalized radius of the CF can be defined as $u = r_{met}/r_{CF}$. The current- and temperature-dependant variation of $u$ is described by

$$u(I_{mem}, T_{amb}) = \exp\left[-\frac{1}{2}W_0\left(\frac{4\pi^2 \kappa r_{CF}^2 (T_{MIT} - T_c)}{I_{mem}^2 \rho_{met}}\right)\right] \quad (3)$$

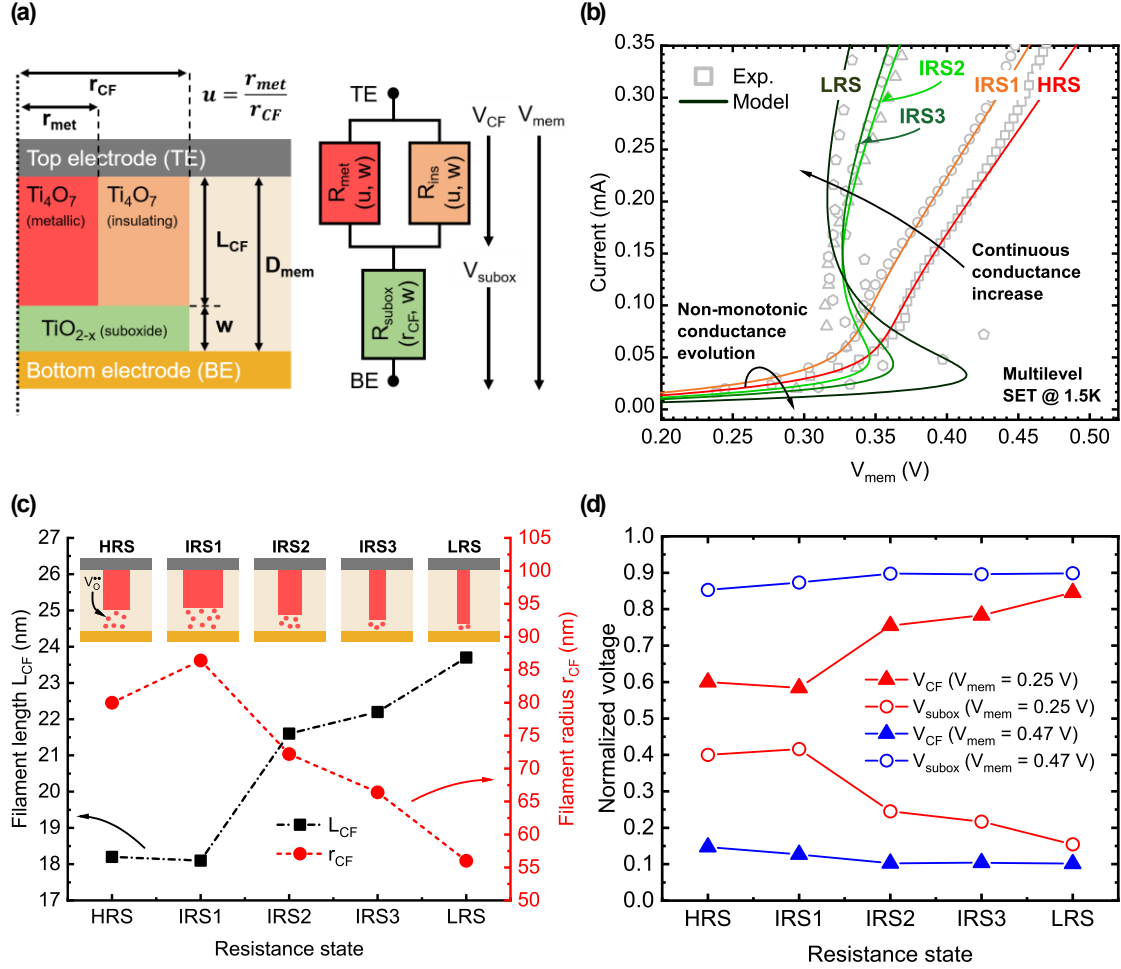

FIG. 4. Analytical model accounting for the non-monotonic conductance evolution of the insulating regime. (a) Schematic representation of the model based on a variable geometry cylindrically symmetric CF, along with the corresponding equivalent electric diagram. (b) Experimental (symbols) *vs.* theoretical (lines) *I-V* curves at 1.5 K for 5 resistance state showing similar behavior: continuous conductance increase in the MR and non-monotonic conductance evolution in the IR. (c) Evolution of the length $L_{CF}$ and radius $r_{CF}$ of the CF as a function of the conductance state of the device, along with schematic representations of the CF geometry evolution. The slight shrinking and spreading of the CF is visible for the IRS1. Red dots represent oxygen vacancies. (d) Contributions of the CF (triangles) and the suboxide (open circles) to the normalized total voltage drop $V_{mem}$ across the device in the insulating regime (red, $V_{mem} = 0.25$ V) and the metallic regime (blue, $V_{mem} = 0.47$ V).

where $W_0(x)$ is the zero-branch Lambert *W* function, $T_{MIT}$ and $T_c$ are the MIT and cryostat temperatures respectively and $I_{mem}$ is the imposed current inside the memristor [19]. The conduction mechanism inside the CF is considered to be ohmic, such as

$$V_{CF}(I_{mem}, u) = R_{CF}(u) \cdot I_{mem} = G_{CF}(u)^{-1} \cdot I_{mem} \quad (4)$$

where $V_{CF}$ is the voltage across the CF and $R_{CF}$ is the total resistance of its two cylindrical phases. In this case, the conductance $G_{CF}(u)$ is described by

$$G_{CF}(u) = \frac{\pi r_{CF}^2}{L_{CF}} \left( \frac{u^2}{\rho_{met}} + \frac{(1-u)^2}{\rho_{ins}} \right). \quad (5)$$

Based on the previous carrier transport analysis results, the gap of length *w* is considered as an oxygen vacancy-rich cylindrically symmetric suboxide of radius $r_{CF}$, whose conduction behavior follows the SCLC model for all resistance states at low current (i.e. in the IR and just after the NDR effect). Using equation (1), the corresponding voltage $V_{subox}$ that depends on the gap $w = D_{mem} - L_{CF}$ is expressed by

$$V_{subox} = \sqrt{\frac{8 I_{mem}(D_{mem} - L_{CF})^3}{9 \pi r_{CF}^2 \varepsilon_0 \varepsilon_r \mu \theta}}. \quad (6)$$

As depicted in the equivalent circuit shown in Fig. 4(a), the total voltage $V_{mem}$ seen by the switching stack during a current-controlled NDR is thus defined by $V_{mem} = V_{CF} + V_{subox}$. Figure 4(b) shows theoretical *I-V* curves calculated with the model compared to experimental data of multilevel SET operations exhibiting NDR effects at 1.5 K. The model successfully reproduce the overall behavior observed in Fig. 2, namely the non-monotonic conductance evolution in the IR and the continuous conductance increase in the MR. Best fitting physical and geometrical parameters are given in Table 1 [28,34,35]. The $L_{CF}$ and $r_{CF}$ values used to fit the different resistance states are visible in Fig. 4(c). Additionally, the contributions of the CF and the suboxide to the normalized total voltage drop $V_{mem}$ across the device are plotted in Fig. 4(d) for each resistance state. The data depicted in Figs. 4(c) and 4(d) suggest that difference in conductance evolution between the IR and the MR originates from the interplay between the geometry and the MIT-driven conduction state of the CF. Indeed, the initial conductance increase from HRS to IRS1 (points A and B in Fig. 2(b)) observable for both conduction regimes of the CF can be explained by an increase of its radius $r_{CF}$, accompanied by a slight decrease of its length $L_{CF}$. This behavior leads to an increase of the section $S = \pi r_{CF}^2$ of the CF, resulting in an increase of the conductance of the device. After the IRS1, one can notice that the subsequent multilevel SET operations induce both an increase of $L_{CF}$ and a decrease of $r_{CF}$. In the IR, the CF is the main contributor to the total voltage drop $V_{mem}$ (see Fig. 4(d)) due to its insulating phase having a higher resistivity than the suboxide. Hence, the growth of the CF provokes a decrease in conductance in this case. However, the opposite behavior is observed in the MR, because of the highly conducting state of the CF in this regime, as illustrated by its negligible contribution to the voltage $V_{mem}$ compared to the suboxide (Fig. 4(d)).

**TABLE 1.** Fitting parameters used for the model

| Parameter | Value | Unit |
|---|---|---|
| $T_c$ | 1.5 | K |
| $T_{MIT}$ | 150 | K |
| $\kappa$ | 4 | W m$^{-1}$ K$^{-1}$ |
| $\varepsilon_r$ | 80 | |
| $\varepsilon_0$ | $8.85 \times 10^{-12}$ | F m$^{-1}$ |
| $\mu$ | $4 \times 10^{-4}$ | cm²/Vs |
| $\theta$ | 0.5 | |
| $\rho_{met}$ | $8 \times 10^{-7}$ | Ohm m |
| $\rho_{ins}$ | $1.1 \times 10^{-2}$ | Ohm m |
| $D_{mem}$ | $31.4 \times 10^{-9}$ | m |

The way $L_{CF}$ and $r_{CF}$ evolves during the multilevel switching stems from a competition between thermal- and field-activated effects [36] exacerbated at low temperature. The widening of the CF between HRS and IRS1 is the result of the strong self-heating-induced temperature gradient between the CF and the rest of the active junction at 1.5 K. This phenomenon activates the diffusion of the oxygen vacancies in the vicinity of the TE towards the center of the CF due to the Soret effect [37]. At such low voltage, this results in an increase of the CF radius in the absence of any field-induced drifting towards the BE. However, from IRS1, the increasingly high applied fields force the thermoactivated oxygen vacancies to drift, which leads to a rapid growth of the CF characterised by a decrease of $r_{CF}$. From these elements, we can assume that the increase in conductance observed at the end of the SET procedure in the IR (points C and D in Fig. 2(b)) could be attributed to a new widening of the CF induced by the thermal- and field-assisted accumulation of oxygen vacancies. Similarly, the ICE observed during the multilevel RESET process is in agreement with the expected shrinking and widening of the CF.

In conclusions, we report successful multilevel resistive switching of CMOS-compatible TiN/Al$_2$O$_3$/TiO$_{2-x}$/Ti/TiN memory devices at 1.5 K, exhibiting a Joule-heating-induced MIT at a critical ambient temperature of 110 K. The conductance increase/decrease in the metallic regime is continuous and gradual, contrary to the insulating regime characterized by a non-monotonic conductance evolution. The carrier transport analysis shows that the main conduction mechanisms in the partially formed filament are SCLC and TAT, the latter being only dominant in the metallic regime for low-resistance states. These results represent direct observations of the interplay between the Joule-heating-dependant conducting regimes and the geometry evolution of the filament, which are respectively impacted by the applied field range and the switching process. The implementation in an analytical model of the transport mechanisms, the MIT-based NDR effects and the geometry-driven conductance states of the filament allows to calculate I-V characteristics with conductance evolution behaviors in good agreement with experimental measurements. This approach reveals both longitudinal and radial evolution dynamics of the conductive filament, resulting from a competition between the diffusion and drift phenomena of oxygen vacancies induced by the strong temperature gradient (Soret effect) and the increasingly high applied electric field respectively. Our study in cryogenic temperature conditions of multilevel resistive switching of TiO$_2$-based devices exhibiting MIT sheds new light on the switching mechanism in oxide-based memories, while highlighting the potential of such highly scalable devices for on-chip cryogenic high-density memory.

## ACKNOWLEDGEMENTS

We would like to thank Michael Lacerte, Marc-Antoine Roux and Prof. Michel Pioro-Ladrière from Institut Quantique of Université de Sherbrooke for their valuable help with cryogenic measurements. This work was supported by Natural Sciences and Engineering Research Council of Canada (NSERC). This research was undertaken thanks in part to funding from the Canada First Research Excellence Fund. We acknowledge financial supports from the EU: ERC-2017-COG project IONOS (# GA 773228).

## REFERENCE

[1] F. A. Zwanenburg, A. S. Dzurak, A. Morello, M. Y. Simmons, L. C. L. Hollenberg, G. Klimeck, S. Rogge, S. N. Coppersmith, and M. A. Eriksson, Silicon quantum electronics, Rev. Mod. Phys. **85**, 961 (2013).

[2] M. Veldhorst, C. H. Yang, J. C. C. Hwang, W. Huang, J. P. Dehollain, J. T. Muhonen, S. Simmons, A. Laucht, F. E. Hudson, K. M. Itoh, A. Morello, and A. S. Dzurak, A two-qubit logic gate in silicon, Nature **526**, 410 (2015).

[3] L. Petit, J. M. Boter, H. G. J. Eenink, G. Droulers, M. L. V. Tagliaferri, R. Li, D. P. Franke, K. J. Singh, J. S. Clarke, R. N. Schouten, V. V. Dobrovitski, L. M. K. Vandersypen, and M. Veldhorst, Spin Lifetime and Charge Noise in Hot Silicon Quantum Dot Qubits, Phys. Rev. Lett. **121**, 076801 (2018).

[4] C. H. Yang, R. C. C. Leon, J. C. C. Hwang, A. Saraiva, T. Tanttu, W. Huang, J. Camirand Lemyre, K. W. Chan, K. Y. Tan, F. E. Hudson, K. M. Itoh, A. Morello, M. Pioro-Ladrière, A. Laucht, and A. S. Dzurak, Operation of a silicon quantum processor unit cell above one kelvin, Nature **580**, 350 (2020).

[5] R. Li, L. Petit, D. P. Franke, J. P. Dehollain, J. Helsen, M. Steudtner, N. K. Thomas, Z. R. Yoscovits, K. J. Singh, S. Wehner, L. M. K. Vandersypen, J. S. Clarke, and M. Veldhorst, A crossbar network for silicon quantum dot qubits, Sci. Adv. **4**, eaar3960 (2018).

[6] J. P. G. van Dijk, E. Charbon, and F. Sebastiano, The electronic interface for quantum processors, Microprocess. Microsyst. **66**, 90 (2019).

[7] J. M. Hornibrook, J. I. Colless, I. D. Conway Lamb, S. J. Pauka, H. Lu, A. C. Gossard, J. D. Watson, G. C. Gardner, S. Fallahi, M. J. Manfra, and D. J. Reilly, Cryogenic Control Architecture for Large-Scale Quantum Computing, Phys. Rev. Appl. **3**, 024010 (2015).

[8] B. Patra, R. M. Incandela, J. P. G. van Dijk, H. A. R. Homulle, L. Song, M. Shahmohammadi, R. B. Staszewski, A. Vladimirescu, M. Babaie, F. Sebastiano, and E. Charbon, Cryo-CMOS Circuits and Systems for Quantum Computing Applications, IEEE J. Solid-State Circuits **53**, 309 (2018).

[9] P. Galy, J. Camirand Lemyre, P. Lemieux, F. Arnaud, D. Drouin, and M. Pioro-Ladriere, Cryogenic Temperature Characterization of a 28-nm FD-SOI Dedicated Structure for Advanced CMOS and Quantum Technologies Co-Integration, IEEE J. Electron Devices Soc. **6**, 594 (2018).

[10] S. Bonen, U. Alakusu, Y. Duan, M. J. Gong, M. S. Dadash, L. Lucci, D. R. Daughton, G. C. Adam, S. Iordanescu, M. Pasteanu, I. Giangu, H. Jia, L. E. Gutierrez, W. T. Chen, N. Messaoudi, D. Harame, A. Muller, R. R. Mansour, P. Asbeck, and S. P. Voinigescu, Cryogenic Characterization of 22nm FDSOI CMOS Technology for Quantum Computing ICs, IEEE Electron Device Lett. 1 (2018).

[11] J.-H. Bae, J.-W. Back, M.-W. Kwon, J. H. Seo, K. Yoo, S. Y. Woo, K. Park, B.-G. Park, and J.-H. Lee, Characterization of a Capacitorless DRAM Cell for Cryogenic Memory Applications, IEEE Electron Device Lett. **40**, 1614 (2019).

[12] F. Wang, T. Vogelsang, B. Haukness, and S. C. Magee, DRAM Retention at Cryogenic Temperatures, in *2018 IEEE Int. Mem. Work.* (IEEE, 2018), pp. 1–4.

[13] S. Slesazeck and T. Mikolajick, Nanoscale resistive switching memory devices: a review, Nanotechnology **30**, 352003 (2019).

[14] F. Zahoor, T. Z. Azni Zulkifli, and F. A. Khanday, Resistive Random Access Memory (RRAM): an Overview of Materials, Switching Mechanism, Performance, Multilevel Cell (mlc) Storage, Modeling, and Applications, Nanoscale Res. Lett. **15**, 90 (2020).

[15] V. Gupta, S. Kapur, S. Saurabh, and A. Grover, Resistive Random Access Memory: A Review of Device Challenges, IETE Tech. Rev. 1 (2019).

[16] R. Fang, W. Chen, L. Gao, W. Yu, and S. Yu, Low-Temperature Characteristics of HfOx-Based Resistive Random Access Memory, IEEE Electron Device Lett. **36**, 567 (2015).

[17] S. Blonkowski and T. Cabout, Bipolar resistive switching from liquid helium to room temperature, J. Phys. D. Appl. Phys. **48**, 345101 (2015).

[18] C. Vaca, M. B. Gonzalez, H. Castan, H. Garcia, S. Duenas, F. Campabadal, E. Miranda, and L. A. Bailon, Study From Cryogenic to High Temperatures of the High- and Low-Resistance-State Currents of ReRAM Ni–HfO2–Si Capacitors, IEEE Trans. Electron Devices **63**, 1877 (2016).

[19] M. D. Pickett, J. Borghetti, J. J. Yang, G. Medeiros-Ribeiro, and R. S. Williams, Coexistence of Memristance and Negative Differential Resistance in a Nanoscale Metal-Oxide-Metal System, Adv. Mater. **23**, 1730 (2011).

[20] H. S. Alagoz, K. H. Chow, and J. Jung, Low-temperature coexistence of memory and threshold switchings in Pt/TiOx/Pt crossbar arrays, Appl. Phys. Lett. **114**, 163502 (2019).

[21] Y. Zhang, N. Deng, H. Wu, Z. Yu, J. Zhang, and H. Qian, Metallic to hopping conduction transition in Ta2O5−x/TaOy resistive switching device, Appl. Phys. Lett. **105**, 063508 (2014).

[22] V. A. Voronkovskii, V. S. Aliev, A. K. Gerasimova, and D. R. Islamov, Conduction mechanisms of TaN/HfOx/Ni memristors, Mater. Res. Express **6**, 076411 (2019).

[23] X.-D. Huang, Y. Li, H.-Y. Li, K.-H. Xue, X. Wang, and X.-S. Miao, Forming-Free, Fast, Uniform, and High Endurance Resistive Switching From Cryogenic to High Temperatures in W/AlOx/Al2O3/Pt Bilayer Memristor, IEEE Electron Device Lett. **41**, 549 (2020).

[24] N. Andreeva, A. Ivanov, and A. Petrov, Multilevel resistive switching in TiO2/Al2O3 bilayers at low temperature, AIP Adv. **8**, 025208 (2018).

[25] Y. Beilliard, F. Paquette, F. Brousseau, S. Ecoffey, F. Alibart, and D. Drouin, Investigation of resistive switching and transport mechanisms of Al2O3/TiO2−x memristors under cryogenic conditions (1.5 K), AIP Adv. **10**, 025305 (2020).

[26] H.-Y. Chen, S. Brivio, C.-C. Chang, J. Frascaroli, T.-H. Hou, B. Hudec, M. Liu, H. Lv, G. Molas, J. Sohn, S. Spiga, V. M. Teja, E. Vianello, and H.-S. P. Wong, Resistive random access memory (RRAM) technology: From material, device, selector, 3D integration to bottom-up fabrication, J. Electroceramics **39**, 21 (2017).

[27] E. Gale, TiO2-based memristors and ReRAM: materials, mechanisms and models (a review), Semicond. Sci. Technol. **29**, 104004 (2014).

[28] R. F. Bartholomew and D. R. Frankl, Electrical Properties of Some Titanium Oxides, Phys. Rev. **187**, 828 (1969).

[29] A. D. Inglis, Y. Le Page, P. Strobel, and C. M. Hurd, Electrical conductance of crystalline TinO2n-1 for n=4-9, J. Phys. C Solid State Phys. **16**, 317 (1983).

[30] E. Lim and R. Ismail, Conduction Mechanism of Valence Change Resistive Switching Memory: A Survey, Electronics **4**, 586 (2015).

[31] W. Kao, K. C.; Hwang, Electrical Transport in Solids, *Electrical Transport in Solids* (Pergamon, Oxford, New York, 1981).

[32] M. P. Houng, Y. H. Wang, and W. J. Chang, Current transport mechanism in trapped oxides: A generalized trap-assisted tunneling model, J. Appl. Phys. **86**, 1488 (1999).

[33] J. Pascual, J. Camassel, and H. Mathieu, Fine structure in the intrinsic absorption edge of TiO2, Phys. Rev. B **18**, 5606 (1978).

[34] S.-M. Lee, D. G. Cahill, and T. H. Allen, Thermal conductivity of sputtered oxide films, Phys. Rev. B **52**, 253 (1995).

[35] M. D. Stamate, On the dielectric properties of dc magnetron TiO2 thin films, Appl. Surf. Sci. **218**, 318 (2003).

[36] D. Ielmini, Resistive switching memories based on metal

oxides: mechanisms, reliability and scaling, Semicond. Sci. Technol. **31**, 063002 (2016).

[37] D. B. Strukov, F. Alibart, and R. Stanley Williams, Thermophoresis/diffusion as a plausible mechanism for unipolar resistive switching in metal–oxide–metal memristors, Appl. Phys. A **107**, 509 (2012).